\documentclass[a4paper]{svmult}
\usepackage{times}
\usepackage[dvips]{graphicx,epsfig}
\usepackage{amsmath,amsfonts,amssymb}
\usepackage{cite,url}

\begin{document}





\newcommand{\be}{\begin{equation}} \newcommand{\ee}{\end{equation}}
\newcommand{\ba}{\begin{eqnarray}} \newcommand{\ea}{\end{eqnarray}}
\newcommand{\uu}{u({\bf p},s)} \newcommand{\ubar}{\overline{u}({\bf
    p'},s')} \newcommand{\nk}{{\bf k}} \newcommand{\nq}{{\bf q}}
\newcommand{\np}{{\bf p}} \newcommand{\nh}{{\bf h}}
\newcommand{\nO}{{\bf 0}} \newcommand{\nkappa}{\mbox{\boldmath
    $\kappa$}} \newcommand{\nx}{{\bf x}}
\newcommand{\neta}{\mbox{\boldmath $\eta$}} 
\newcommand{\nsigma}{\mbox{\boldmath $\sigma$}} 
\newcommand{\ny}{{\bf y}}
\newcommand{\nR}{{\bf R}} \newcommand{\sigvec}{\mbox{\boldmath
    $\sigma$}} \newcommand{\etavec}{\mbox{\boldmath $\eta$}}
\newcommand{\tauvec}{\mbox{\boldmath $\tau$}}
\newcommand{\kappavec}{\mbox{\boldmath $\kappa$}}
\newcommand{\kvec}{\mbox{\boldmath $\zeta$}}
\newcommand{\avec}{\mbox{\boldmath $A$}}
\newcommand{\bvec}{\mbox{\boldmath $B$}}
\newcommand{\gammavec}{\mbox{\boldmath $\gamma$}}
\newcommand{\Gammavec}{\mbox{\boldmath $\Gamma$}}
\newcommand{\jvec}{\mbox{\boldmath $J$}}
\newcommand{\hvec}{\mbox{\boldmath $h$}} \newcommand{\bd}[1]{
  \mbox{\boldmath $#1$} } \newcommand{\sla}[1]{#1 \!\!\!\!\!\!\slash }
\def\be{\begin{equation}}
\def\ee{\end{equation}}
\def\bea{\begin{eqnarray}}
\def\eea{\end{eqnarray}}
\def\bear{\begin{array}}
\def\ear{\end{array}}
\def\bfig{\begin{figure}}
\def\efig{\end{figure}}
\def\bcen{\begin{center}}
\def\ecen{\end{center}}
\def\raw{\rightarrow}
\def\bra#1{\left\langle #1\right|}
\def\ket#1{\left| #1\right\rangle}
\def\Bra#1{\bigl\langle #1\bigr|}
\def\Ket#1{\bigl| #1\bigr\rangle}
\def\vp{\mathbf{p}}
\def\vk{\mathbf{k}}
\def\vv{\mathbf{v}}
\def\vq{\mathbf{q}}
\def\mpi{m_{\pi}}
\def\la{\label}
\def\chic{\scriptscriptstyle}
\def\S{\scriptstyle}
\def\D{\displaystyle}
\def\slash{\!\!\! /} 
\def\bpi{\boldsymbol\pi}
\def\btau{\boldsymbol\tau}
\def\bsigma{\boldsymbol\sigma}
\def\bphi{\boldsymbol\phi}
\def\brho{\boldsymbol\rho}
\def\bnu{\boldsymbol\nu}
\def\bmu{\boldsymbol\mu}
\def\bkappa{\boldsymbol\kappa}
\def\boeta{\boldsymbol\eta}

\def\three_j(#1,#2,#3,#4,#5,#6){\pmatrix{#1 & #2 & #3\cr
                                         #4 & #5 & #6\cr}}

\def\qqq{\end{document}}
\def\pmb#1{\setbox0=\hbox{$#1$}%
\kern-.025em\copy0\kern-\wd0
\kern.05em\copy0\kern-\wd0
\kern-.025em\raise.0433em\box0 }
\def\b{\pmb}

\def\xara(#1,#2,#3,#4){\left(\matrix{#1 & #2\cr #3 & #4\cr}\right)}
\def\Fn{J_{(n)}}
\def\Fs{J_{(s)}}
\def\Hn{H_{(n)}}
\def\Hs{H_{(s)}}
\def\thru#1{\mathrel{\mathop{#1\!\!\!/}}}
\def\CN{\cal N}
\def\w{\omega}
\def\W{\Omega}
\def\six_j(#1,#2,#3,#4,#5,#6){\left\{\matrix{#1 & #2 & #3\cr
                                         #4 & #5 & #6\cr}\right\}}
\def\nine_j(#1,#2,#3,#4,#5,#6,#7,#8,#9){\left\{\matrix{#1 & #2 & #3\cr
                                        #4 & #5 & #6\cr
                                         #7 & #8 & #9\cr}\right\}}
\def\W{\Omega}
\def\pd#1#2{{\partial #1\over \partial #2}}
\def\v#1{ {\bf #1} }
\def\Ener(#1,#2){ \sqrt{{#1}^2+{#2}^2} }
\def\c#1{ {\cal #1}}%

\def\overlay#1#2{\setbox0=\hbox{$#1$}\setbox1=\hbox to \wd0{\hss$#2$\hss}#1%
\hskip -1\wd0\copy1}
\newcommand{\xslash}[1]{\overlay{#1}{/}}
\newcommand{\undsim}[1]{\olay{#1}{\sim}}

\def\bold#1{\setbox0=\hbox{$#1$}%
      \kern-.025em\copy0\kern-\wd0
      \kern.05em\copy0\kern-\wd0
      \kern-.025em\raise.0433em\box0 }
\def\tr{\, \hbox{tr} \, }
\def\Tr{\, \hbox{Tr} \, }
\def\sgn{\, \hbox{sgn} }
\def\bra{\langle}
\def\ket{\rangle}
\def\shalf{\,1/2\,}
\def\half{\, {1 \over 2} \,}
\def\gsim{\displaystyle\mathop{>}_{\sim}}
\def\lsim{\displaystyle\mathop{<}_{\sim}}
\def\rd{\partial}
\def\c{\hbox{c}}
\def\s{\hbox{s}}

\def\S11{S_{11}(1535)}
\def\etaNN{\eta NN^*}
\def\geta{g_{\eta NN^*}}

\def\tdotr{\, \vec \tau \cdot \hat r \, }
\def\fpi{f_\pi}

\renewcommand{\thefootnote}{\fnsymbol{footnote}}
\def\footnoterule{\kern-3pt \hrule width \hsize \kern2.6pt}

\newcommand{\nne}{{\bf e}}
\newcommand{\im}{{\rm Im}\,}
\newcommand{\re}{{\rm Re}\,}
\newcommand{\qbar}{\not{\!Q}}
\newcommand{\pbar}{\not{\!P}}
\newcommand{\pdbar}{\not{\!P}_{\!\Delta}}

%
\newcommand{\md}{m_\Delta}

\newcommand{\na}{{\bf      a}}  
\newcommand{\nb}{{\bf      b}}     
\newcommand{\nj}{{\bf      j}}
\newcommand{\nr}{{\bf      r}}         
\newcommand{\ns}{{\bf      s}}
\newcommand{\nv}{{\bf      v}}
\newcommand{\nw}{{\bf      w}}
\newcommand{\nA}{{\bf      A}}       
\newcommand{\nB}{{\bf      B}}
\newcommand{\nC}{{\bf      C}}
\newcommand{\nJ}{{\bf      J}}
\newcommand{\nM}{{\bf      M}}       
\newcommand{\nP}{{\bf      P}}       
\newcommand{\nS}{{\bf      S}}      
\newcommand{\nX}{{\bf      X}}
\newcommand{\nY}{{\bf      Y}}          
\newcommand{\hp}{{\bf \hat{p}}}
\newcommand{\hr}{{\bf \hat{r}}} 
\newcommand{\hx}{{\bf \hat{x}}}

\title*{Electron and neutrino scattering
in the $\Delta$-resonance region and beyond}

\author{\underline{{Maria B. Barbaro}}\inst{1}
\and {J.E. Amaro}\inst{2}
\and {J.A. Caballero}\inst{3}
\and {C. Maieron}\inst{4}
}

\titlerunning{Electron and neutrino scattering
in the $\Delta$-resonance region and beyond}
\authorrunning{{Maria B. Barbaro} {\em et al.}}


\institute{
{Universit\`a di Torino and INFN, Sezione di Torino, Italy}
\and
{Departamento de F\'{\i}sica At\'{o}mica, Molecular y Nuclear, 
Universidad de Granada,Spain}
\and
{Departamento de F\'{\i}sica At\'{o}mica, Molecular y Nuclear,
Universidad de Sevilla, Spain}
\and
{Universit\`a di Lecce and INFN, Sezione di Lecce, Italy}
}

\maketitle

\begin{abstract}
We present a unified relativistic approach to inclusive electron 
scattering based on the relativistic Fermi gas model and on a phenomenological
extension of it which accounts for the superscaling behaviour of $(e,e')$
data. We present results in the $\Delta$ resonance region and in the
highly inelastic domain and show some application to neutrino scattering.
\end{abstract}

\section{Introduction}

Electron scattering off complex nuclei is an ideal testing ground for 
modeling neutrino-nucleus cross sections, whose accurate prediction is 
necessary for the analysis of on-going experimental studies of 
neutrino oscillations at GeV energies,
presently being pursued in the MiniBooNE and K2K/T2K 
experiments~\cite{AguilarArevalo:2007it,Ahn:2006zz}.

Indeed inclusive electron $(e,e')$ and charged-current (CC) neutrino 
$(\nu_l,l^-)$ are closely related processes:
in first Born approximation they involve the response of the 
nuclear system to a virtual boson, a photon or a $W^+$, probing the
electromagnetic and weak nuclear currents, respectively.

Since the typical neutrino energies in oscillation experiments are of a 
few GeV, we will focus our attention on this kinematical domain.
In this case the inclusive $(l,l')$
cross section shows a pronounced peak, the so-called quasielastic peak
(QEP), at an energy transfer 
$\omega \sim \sqrt{q^2+m_N^2}-m_N$, corresponding to the
quasi-free interaction with the individual nucleons in the nucleus 
(here $m_N=$ nucleon mass). 
For high values of the momentum transfer $q=|{\bf q}|$ and higher energy 
loss it is possible to  produce real pions and the cross section
shows a second peak dominated by the resonant production
of a $\Delta(1232)$ at
$\omega \sim \sqrt{q^2+m_\Delta^2}-m_N$, where $m_\Delta$ is the $\Delta$
mass.
The width of these peaks is related to the Fermi momentum of the 
nucleons inside the nucleus and, in the case of the $\Delta$-peak,
also to the decay width of the $\Delta$ in nuclear matter.
Hence for a high enough value of $q$, these two peaks actually 
overlap and cannot be separated in inclusive experiments.
At higher energy transfer the so-called second resonance region
is found, where the $N^*(1440)P_{11}$ (Roper), $N^*(1520)D_{13}$
and $N^*(1535)S_{11}$ resonances are excited, evolving, at very
high energies, into the Deep Inelastic Scattering (DIS) regime.
  
Understanding the above spectrum in a unified framework has been
the aim of some recent studies, which will be briefly
summarized in this contribution. We shall mainly address the electron
scattering problem, where more data are available, and mention some 
applications to neutrino reactions.

\section{The $\Delta$ resonance region}

Since the kinematical domain we are exploring involves energy and momentum 
transfers of the order of (or higher than) the nucleon mass, the traditional 
non-relativistic approach is bound to fail and a relativistic approach
to the problem is required. 

A fully relativistic treatment of the nuclear many-body system is a
longstanding and extremely difficult task that is best pursued in 
special frameworks like the Relativistic Fermi Gas (RFG) model, where 
basic symmetries like Lorentz covariance and gauge invariance are exactly 
respected.

Hence we assume as a starting point an extension of the 
RFG model, which has been widely employed in the QEP, to the inelastic region. In this model, the virtual
boson is absorbed by an on-shell nucleon described by a Dirac spinor
$u(\nh,s_h)$, with energy $\overline{E}_\nh=\sqrt{\nh^2+m_N^2}$.

The basic ingredient of the calculation
is the hadronic tensor ${\cal W}^{\mu\nu}$, which 
contains all the information on the nuclear structure and dynamics and  yields 
the $(e,e')$ 
differential cross section with respect to the lepton final energy 
$\varepsilon_f$ and solid
angle $\Omega_f$ according to the follwing relation
\begin{equation}
\frac{d\sigma}{d\Omega_f d\varepsilon_f} = \frac{2\alpha^2}{Q^4}
\frac{\varepsilon_f}{\varepsilon_i} \eta_{\mu\nu} {\cal W}^{\mu\nu}~,
\label{eq:1}
\end{equation}
where $\alpha$ is the fine structure constant, 
$\varepsilon_i$ the lepton initial energy,
$\eta_{\mu\nu}$ the leptonic tensor and 
$Q_\mu=(\omega,\nq)$ the four-momentum transfer.

For the excitation of a stable resonance $N^*$ of mass $m^*$ the RFG 
hadronic tensor turns out to be~\cite{Barbaro:2003ie}
\be
{\cal W}_{*,RFG}^{\mu\nu}(q, \omega, m^*) = 
\frac{3{\mathcal N}}{4\kappa m_{\chic{N}} \eta_{\chic{F}}^3}
\xi_{\chic{F}}\,
\Theta (1 - \psi^{*2}) (1 - \psi^{*2})  
U_*^{\mu\nu}(q,\omega, m^*) 
\label{eq:2}
\ee
where ${\mathcal N}$ is the number of nucleons involved in the reaction,
$\kappa=q/2 m_N$ and $\eta_F=\sqrt{\xi_F(\xi_F+2)}=k_F/m_N$ 
are the dimensionless transferred 
and Fermi momentum, respectively, 
and the ``single-nucleon''
tensor $U_*^{\mu\nu}$ embeds the information about the $N-N^*$ transition
~\footnote{Actually in a fully relativistic 
framework the single nucleon physics cannot be exactly disentagled from 
the many-body part of the problem and the tensor $U_*^{\mu\nu}$ contains
corrections due to the medium (see, e.g., ref.~\cite{AlvarezRuso:2000bx}).}
.

In analogy with the physics of the quasi-elastic
peak~\cite{Alberico:1988bv}, a {\em scaling variable} 
$\psi^*$, defined as follows
\ba
\psi^* &\equiv& \psi(\kappa,\lambda; m^*) = \pm
\sqrt{ \frac{1}{\xi_F}
\left[ \kappa \sqrt{\frac{1}{\tau}+\rho(m^*)^2}
-\lambda\rho(m^*) - 1\right] }\ , 
\label{eq:3}
\ea
with  $\lambda=\omega/2m_N$ and $\tau=\kappa^2-\lambda^2$, has been introduced.
The quantity
\ba
\rho(m^*) = 1 + \frac{1}{4\tau} \left(m^{*2}/m_N^2-1\right)
\ ,
\label{eq:4}
\ea
measures the inelasticity of the elementary process and reduces to 1 in the
quasielastic limit $m^*=m_N$.
The physical meaning of the scaling variable $\psi^*$ is the following:
$\xi_F \psi^{*2}$ represents the minimum kinetic energy 
required to transform a nucleon inside the nucleus into a resonance 
$N^*$ when hit by a photon of energy $\lambda$ and momentum $\kappa$.
In terms of the scaling variable the RFG response region 
associated to a specific resonance is $-1\leq\psi^*\leq 1$ and the
peak position corresponds to $\psi^*=0$.

A realistic model for the resonance requires the inclusion of the 
decay width $\Gamma$ in the hadronic tensor. This
can be computed from the tensor ${\cal W}_*^{\mu\nu}(q,\omega,W)$
for a stable $N^*$ with mass $W$  by a convolution
\begin{equation}
{\cal W}^{\mu\nu}_\Gamma(q,\omega)= \int_{W_{min}}^{W_{max}} 
\frac{1}{\pi}\frac{\Gamma(W)/2}{(W-m^*)^2+\Gamma(W)^2/4}
{\cal W}_*^{\mu\nu}(q,\omega,W)dW \ ,
\label{eq:7}
\end{equation}
where the integration interval goes from threshold
to the maximum value allowed in the Fermi gas model.

The above expressions are independent of the specific transition under
consideration: they can be used to describe the $\Delta$, the 
$N^*(1440)$ or any other resonance
region~\cite{Amaro:1999be,AlvarezRuso:2003gj} 
(after modifying $m^*$ and $\Gamma(W)$ accordingly) 
as well as the quasielastic peak (in the $\Gamma \raw 0$ limit setting
$m^*=m_{\chic{N}}$)~\cite{Donnelly:1991qy}.
\begin{figure}[htp]
\begin{center}
  \includegraphics[scale=0.6]{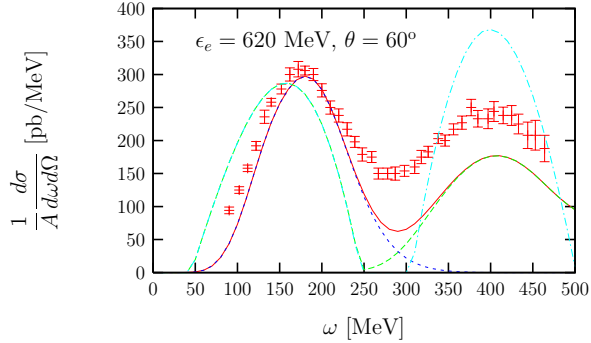}
\caption{Inclusive cross section per nucleon
from $^{12}$C.
Llight blue: RFG without $\Delta$ width; green: 
RFG including the finite  $\Delta$ width; blue: 
PWIA for the quasi-elastic peak;
red: hybrid model obtained by adding the PWIA cross section 
in the quasi-elastic 
peak and RFG cross section in the $\Delta$-peak. 
Experimental data are taken from \cite{Bar83}. 
}
\end{center}
\end{figure}

As already mentioned the specific transition form factors are hidden in the 
tensor $U_*^{\mu\nu}$.
Concerning the $\Delta$ resonance, a relativistic calculation of the 
$(e,e')$ response in this
region must use the full $N\to\Delta$ vertex, which includes
the magnetic (M1), electric (E2) and Coulomb (C2) excitation amplitudes.
Although for years an important program has been pursued to determine 
more accurately 
the quadrupole $C2$ and $E2$ amplitudes in the $\Delta$ region 
(these being small compared with the dominant dipole $M1$
amplitude), our knowledge is still
incomplete and has not been possible to undertake a full analysis
in the  sense of the work by Nozawa and Lee \cite{Noz90}
of the effect of the C2 and E2 form factors in the nuclear
$\Delta$-peak.
Accordingly we use the parameterization~\cite{Weh89}
$
G_{E,M}(Q^2) = G_{E,M}(0) G_E^P(Q^2)/
\sqrt{1-Q^2/(3.5 GeV/c)^2}
$,
namely we assume that the same dependence in $Q^2$ is valid for the
electric and magnetic form factors and that the isobar form factor 
falls off faster than the proton form factor.

In Fig.~1 we show results for the nuclear inclusive cross section 
per nucleon from $^{12}$C compared with the experimental data.
Light-blue lines correspond to the RFG 
for a stable $\Delta$. 
The width is included in the green line result according to eq.~(5) and
produces a broadening  of the
$\Delta$ peak and correspondingly a decrease of the strength. 
As an illustration of how one could improve the model in the 
quasi-elastic-peak region, we also show with blue lines the 
quasi-elastic cross section computed with the 
semirelativistic PWIA model of ref.
\cite{Ama96b}, which includes the momentum distribution of 
the finite-sized nucleus,
producing the ``tails'' of the cross section, and the 
binding energy of the nucleons in the nucleus, yielding a shift to higher
energies. 
Finally, we show with red lines the results computed with a hybrid 
model in which we add the PWIA cross section for the quasi-elastic 
contribution to the RFG result for the $\Delta$ contribution.

As we can see in Fig.~1, our results are  below the data 
in the dip and $\Delta$ region. This was expected because other
contributions coming mainly from two-nucleon emission and 
non-resonant pion production (not included in our model)
also enter here.

In Ref.~\cite{Amaro:1999be} several relativistic effects and ingredients of the
calculation were analyzed in detail. Here we just like to stress two basic 
findings.
First, 
from a comparison between different Lagrangians in
the treatment of the $\Delta$ excitation, it emerges that the Peccei
Lagrangian~\cite{Pec69}, employed in the pioneering calculation by Moniz \cite{Mon69}, is only appropriate 
for computing the transverse response for low momentum transfer, but in the longitudinal channel the full
vertex~\cite{Jon73} 
\begin{equation}
\langle \Delta | j_\mu | N\rangle
= \overline{u}_\Delta^\beta
\left( C_1\Gamma^1_{\mu\beta}
+C_2\Gamma^2_{\mu\beta}
+C_3\Gamma^3_{\mu\beta}\right)
u \ ,
\label{matrix}
\end{equation}
(where $u_\Delta^\beta$ is the Rarita-Schwinger spinor describing a spin 3/2 particle)
should be retained: indeed the Peccei Lagrangian, corresponding to the
first term $\Gamma^1$ only, gives an unreasonably large longitudinal response. 
Second, we have found a large sensitivity of the longitudinal
response to the inclusion of the Coulomb form factor of the isobar, 
especially for high
$q$, a fact that could be of importance for investigations
of the longitudinal nuclear response. 
On the other hand, both longitudinal and transverse responses are 
found to be insensitive to the quadrupole E2 form factor.  

Let us now briefly comment on the second resonance region,
namely where resonances heavier than the $\Delta$ are found.
Among these, the $N^*(1440)P_{11}$ (the so-called Roper resonance),
occurring just above the $\Delta$(1232), is particularly intersting, since
it can be viewed as a radial excitation of a three-quark nucleon state,
analog of the breathing mode of the nucleus, hence carrying informations on
the nucleon's compressibility.
A first step towards a theoretical
description of the nuclear response functions in the Roper resonance region
was made in ref.\cite{AlvarezRuso:2003gj}, where the corresponding RFG 
responses were calculated within several different theoretical models
for the electroproduction amplitudes. In particular it was shown that,
although the experimental information 
was still insufficient to allow a stringent test of the various theories,
the longitudinal response $R_{\chic{L}}$ associated with the Roper can be
large compared with the contribution arising from the $\Delta$ near the
light-cone and that that the impact of the Roper on the Coulomb sum rule 
can be significant.

\section{The highly inelastic region}

To study what we call the ``highly inelastic'' domain, namely the region
beyond the resonances, we assume that the final state can be 
described in terms of a recoiling nuclear state plus a highly inelastic 
state of mass $m^*$.
In this case the RFG hadronic tensor can be written as~\cite{Barbaro:2003ie}
\begin{eqnarray} 
{\cal W}^{\mu\nu}_{inel,RFG}(q,\omega) 
&=&
\frac{3{\mathcal N}\tau \xi_{\chic{F}}}{2\kappa m_{\chic{N}} \eta_{\chic{F}}^3}
\,
\int_{\rho_1}^{\rho_2}
d\rho
\Theta (1 - \psi^{*2}) (1 - \psi^{*2})  
U_{inel}^{\mu\nu}(q,\omega, m^*) 
\end{eqnarray}
where $\rho_1$ and $\rho_2$ are the kinematical boundaries at fixed
$q$ and $\omega$.
Thus for each value of $\rho$ (and hence $m^*$) a ``peak'' can be
identified, corresponding to the region 
$-1\leq\psi^*\leq 1$, 
centered at $\psi^*=0$, whose width
is a function that grows with $q$ and decreases with $m^*$.
The single-nucleon inelastic hadronic tensor
$U^{\mu\nu}_{inel}$ can be parameterized in terms of
two structure functions, $w_1$ and $w_2$ (see ref.~\cite{Barbaro:2003ie}).
In computing it
we employ phenomenological fits of the single-nucleon
inelastic structure functions measured in DIS experiments.
Among the various parameterizations which can be found in the
literature 
here we adopt the Bodek {\it et al.} fit
of~\cite{Bodek}, 
which describes both the deep inelastic and
resonance regions.

Once the inelastic RFG modeling is in hand, we can 
go beyond this simple non-interacting model,
still retaining its relativistic content.
This we do by exploiting the so-called
``superscaling'' behavior of inclusive lepton-nucleus reactions.
We shall see that this has a significant impact on the nuclear responses at
high inelasticity.

In order to introduce the concept of superscaling, we observe that
Eq.~(\ref{eq:2}) can be recast in the form
\be
{\cal W}_{*,RFG}^{\mu\nu}(q, \omega, m^*) = 
f_{RFG}(\psi^*)\times
G^{\mu\nu}(q,\omega, m^*) \,,
\label{eq:5}
\ee
where
\be
f_{RFG}(\psi^*)=\frac{3}{4}\Theta (1 - \psi^{*2}) (1 - \psi^{*2})  
\label{eq:6}
\ee
is the RFG {\em superscaling function}: hence by dividing the RFG nuclear 
tensor by an appropriate tensor $G^{\mu\nu}$ - whose definition follows
immediately from Eqs.~(\ref{eq:2}) and (\ref{eq:5}) - a universal function
$f_{RFG}$ is obtained, which does not depend on the three variables $q$,
$\omega$ and $k_F$ independently, but only upon one specific combination, 
namely the scaling variable (\ref{eq:3}).

As a consequence the RFG model predicts that if the inclusive
cross section is divided by an appropriate function and plotted versus the 
corresponding scaling variable, no dependence on the momentum transfer $q$ 
(scaling of the first kind) nor on the target nucleus, specified by
the fermi momentum $k_F$ (scaling of the second kind), is found.
The simultaneous occurrence of the two kinds of scaling is known as 
{\em superscaling}.

In the quasielastic peak domain $(m^*=m_N)$, 
superscaling has been widely tested against the $(e,e')$ data 
in Ref.~\cite{Donnelly:1998xg}, where it was shown that both kinds of scaling 
are fulfilled with a good  degree of accuracy 
in the region $\psi<0$ providing the momentum transfer is not too low. 
The scaling analysis of the separated longitudinal and transverse responses
has also proved that the scaling violations observed in the $\psi>0$ region
mainly reside in the transverse channel, whereas the longitudinal data
do scale rather well in the whole QE region. 

However the shape of the experimental QE longitudinal superscaling function
differs from the RFG parabolic one,
extending outside the region $-1<\psi<1$ 
and, most importantly, displaying an asymmetry around the QEP with a
pronounced tail in the $\psi>0$ region. 
The asymmetry of the superscaling function has been and still is
the object of many investigations~\cite{Caballero:2007tz,
Amaro:2006if,Amaro:2006tf,Caballero:2005sj,Amaro:2005dn}
and represents a stringent constraint
that any realistic nuclear model aiming to describe $(l,l')$ reactions should
fulfill. 

An expression for a phenomenological QE
longitudinal scaling function, $f_{QE}(\psi)$, 
was obtained by fitting the data~\cite{Jourdan:1996}.
Based on these results, we now make the following hypothesis:
we assume that $f_{QE}(\psi)$ provides a good description of 
$f(\psi^*) = f_L(\psi^*) =  f_T(\psi^*)$
(``scaling of the zeroth kind''), as it implicitly contains the initial-state physics, 
and thus we make, {\em for any $m^*$}, the following substitution:
\be
f_{RFG}(\psi^*) \rightarrow f_{QE}(\psi^*)
\;.
\label{eq:f_univ2}
\ee
The corresponding results are illustrated in Fig.~2, where
the inclusive cross section is shown 
together with the separated QE and inelastic contributions
for a $^{12}$C
target at $E_{inc}= 4.045$ GeV and 
$\theta_e=15^\circ$ (a), $30^\circ$ (b),
$45^\circ$ (c) and $74^\circ$ (d),
The calculation was performed in the Relativistic Fermi Gas 
including a phenomenological energy shift (red lines)
and in the phenomenological extension of it (magenta lines), which we denote
here as ERFG (Extended Relativistic Fermi Gas), 
based on the fit of the quasielastic scaling function $f_{QE}$~\cite{Jourdan:1996}. 
\begin{figure}[htp]
\begin{center}
\vskip 1.cm
\hskip -2.cm
\includegraphics[scale=0.6]{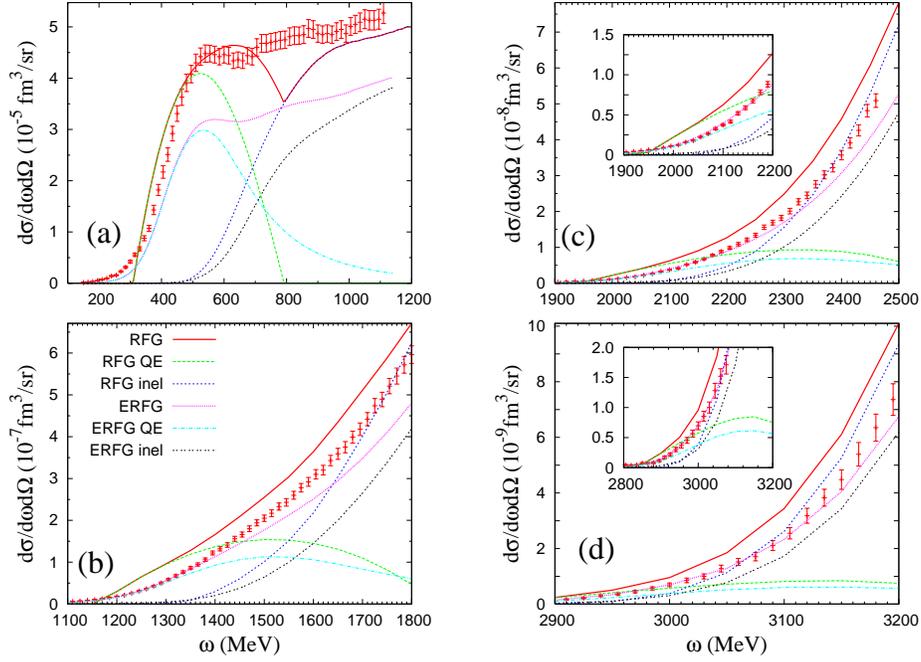}
\caption{Inclusive cross section for electron scattering
from carbon at $E_{inc}= 4.045$ GeV and 
$\theta_e=15^\circ$ (a), $30^\circ$ (b),
$45^\circ$ (c) and $74^\circ$ (d),
versus the energy transfer. Red: RFG; magenta: ERFG. 
The calculation includes an
energy shift $\omega_{shift}=20$ MeV and the separate
QE and inelastic contributions to the cross section are shown (dashed).
Data are from~\protect\cite{Arrington:1998ps}.
}
\end{center}
\end{figure}

We notice that for low scattering angle (a)
the RFG model yields roughly 
the right position and height of the QE peak, but fails to reproduce 
the tails of the peak, 
giving in particular an unobserved dip at $\omega\simeq 800$ MeV.
On the other hand the ERFG, 
while reproducing the data in
the tails better, significantly underestimates the cross section at the peak.
This is related to the fact that
the peak of the phenomenological function $f_{QE}$ is lower than the
corresponding RFG value. 
For higher angles [Figs.~2 (b),(c),(d)] the data lie 
roughly in between the predictions of ERFG (smaller) 
and RFG (larger) models, the former again reproducing the low-$\omega$ 
behavior better.
As a general result we observe that as the scattering angle increases
the range of validity of the ERFG also increases.

An important comment is in order. 
The RFG and ERFG models consider here
the 1p-1h one-body contributions both
for elastic scattering from a nucleon in the nucleus 
and for representations of
the single-nucleon inelastic spectrum, thereby incorporating 
effects from meson production,
excitation of baryon resonances (notably the $\Delta$) and, 
at high excitation energies,
DIS. However, in this 
region and beyond effects arising from reaction mechanisms not 
included here, namely, those
coming from correlations and both 1p-1h and 2p-2h meson-exchange currents 
can be also important~\cite{Amaro:2003yd,Amaro:2002mj,DePace:2003xu} and,
from some preliminary study, they tend to bring the 
total (the present ERFG contributions plus these additional MEC contributions) 
into better agreement with the data. 
Therefore the fact that the ERFG
yields a cross section that is below the data is somehow encouraging, 
since this leaves room for the above-mentioned effects to provide the balance. 

Similar results are obtained for the superscaling function:
in Fig.~3 the total scaling function $f$ is shown as a
function of the QE variable $\psi^\prime$ (the ``prime'' indicating 
the inclusion of a 
phenomenological energy shift, needed to reproduce the QEP position)
for four different nuclei, within the RFG (left panel) and ERFG (right panel)
models, at $E_e=3.595$ GeV and $\theta_e=16^0$;
experimental data are obtained from the measured inclusive cross
sections divided by the single nucleon cross section
and the curves are obtained by dividing the theoretical inclusive
cross section by the same quantity.
A closer inspection of the 
transverse superscaling function $f_T(\psi^\prime)$, performed in 
ref.\cite{Barbaro:2003ie} at the same kinematics, shows that 
the discrepancy between ``data'' and
``theory'' is larger for the transverse case than for the total
scaling functions at this scattering angle
($\theta=16^0$). This indicates that extra contributions should
be added to the nuclear model, going beyond the present one-body
description, and that these must act mainly in the transverse channel.
\begin{figure}[ht]
\begin{center}
\includegraphics[scale=0.6]{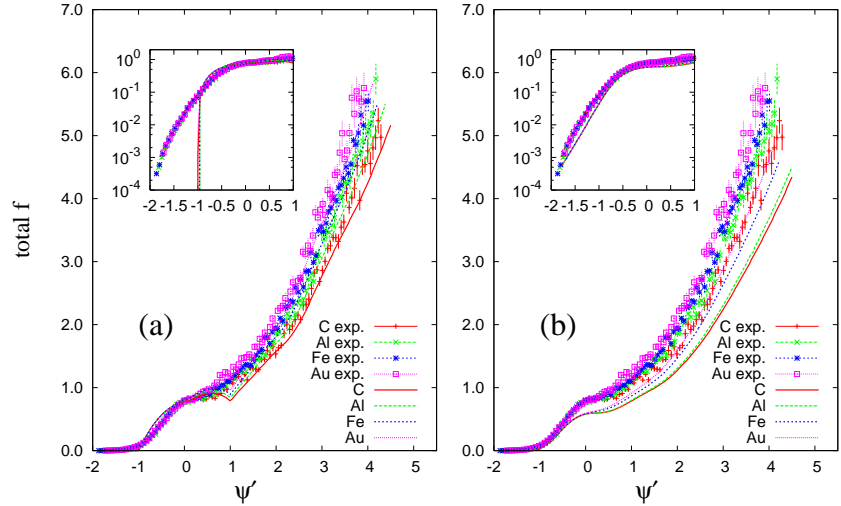}
\caption{Total superscaling functions $f(\psi^\prime)$
for $E_e=3.595$ GeV and $\theta_e=16^\circ$. Theoretical results obtained
within the RFG are shown in panel (a), while the ERFG case is presented
in panel (b).}
\end{center}
\end{figure}

\section{Scaling in the $\Delta$ region and application to neutrino scattering}

In the previous Section the superscaling function $f_{QE}$ extracted
from the quasielastic peak data has been used to calculate the full inelastic
$(e,e')$ spectrum. Two basic assumptions underly this approach: first that
the longitudinal and transverse superscaling functions coincide (0-th
kind scaling) and second that the superscaling function, embodying 
the nuclear initial and final state interactions, is the same in all 
kinematical domains, the latter being characterized only by the structure 
functions $w_1$ and $w_2$.

In Ref.~\cite{Amaro:2004bs}
a different approach to the $\Delta$ resonance region was 
taken.  First the contribution of the $\Delta$ has been
isolated by subtracting from the total experimental cross section 
the quasielastic contribution, reconstructed using the 
superscaling function $f_{QE}$ introduced above.
Next the left-over cross section has been 
divided by the appropriate $N\to\Delta$
single-nucleon cross section and the result has been displayed 
versus the scaling variable $\psi_\Delta$, given by Eq.~(3) 
when $m^*=m_\Delta$.
The results are found to scale quite well~\cite{Amaro:2004bs}
for $\psi_\Delta<0$, suggesting that this procedure
has indeed identified the dominant contributions not only in the QE region,
but also in the $\Delta$ region. Of course for $\psi_\Delta>0$ higher
resonances come into play and this procedure is no longer correct.
The residual scaling function $f^\Delta$, whose fit is plotted in Fig.~4,
while similar to $f_{QE}$, differs in detail:
it is somewhat lower, is shifted slightly and is more spread out over 
a wider range of scaling variable. This is not unexpected, since implicit 
in this approach is the fact that the $\Delta$ brings with it its own width 
and shift.  
A microscopic analysis of $f_\Delta$ is presently being carried 
out~\cite{Chiara}
in order to deconvolute the width from the total response and see whether 
or not the underlying scaling function is indeed the basic $f_{QE}$ 
deduced above. 
\begin{figure}[ht]
\begin{center}
  \includegraphics[scale=0.6]{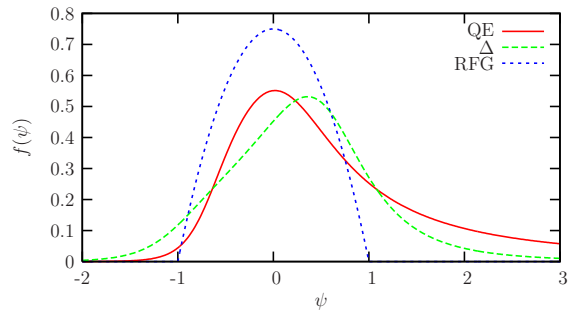}
\caption{Fits of the superscaling function for the 
quasi-elastic (QE) and $\Delta$-resonance regions plotted
versus the corresponding scaling variable $\psi$ and compared to the RFG 
result.}
\end{center}
\end{figure}

As mentioned in the introduction, a major advantage of the approach
above illustrated is that the scaling functions $f_{QE}$ and $f_\Delta$ 
extracted by $(e,e')$ data can be used to predict neutrino-nucleus cross 
sections in both the quasielastic and $\Delta$ regions. 
Indeed, as tested in ref.~\cite{Amaro:2004bs} in a wide range of kinematical
conditions, these phenomenological
functions give {\em by construction} a good description of
the electron scattering data and are in this sense model-independent.
The so-called SuSA (SUperScaling Approximation)
approach amounts to multiply the two superscaling
functions for the corresponding neutrino-nucleon elementary cross section.

Several applications of the SuSA approximation to neutrino and antineutrino
scattering are shown in Ref.~\cite{Amaro:2004bs}.
In Fig.~5 we display, as an example, a comparison of the SuSA and 
RFG results, clearly showing that the relativistic Fermi gas badly
overestimate the cross section in both the QE and $\Delta$ regions as compared 
to the phenomenological model based on superscaling.
\begin{figure}[hbt]
\begin{center}
\hspace*{-1cm}\includegraphics[scale=0.5,clip]{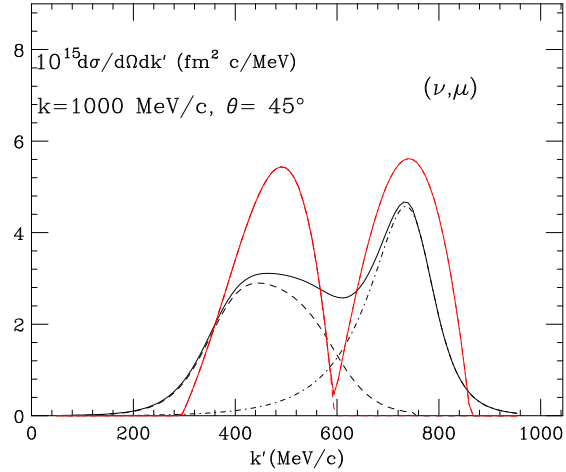}
\parbox{13cm}{\caption[]{Neutrino reaction cross section
for $E_\nu=1$ GeV and $\theta=$ 45 degrees. Red line: RFG result; 
black lines: SuSA result 
(dashed: $\Delta$; dotdashed:quasielastic, solid:total) .
}\label{rfgcomp}} 
\end{center} 
\end{figure}

It is worth mentioning that more fundamental approaches, 
based on relativistic mean field theory
with final state interactions~\cite{Caballero:2005sj,Amaro:2006if}
and on the coherent density fluctuation
model (CDFM)~\cite{Antonov:2007vd,Antonov:2006md} 
and aiming to justify the properties of the 
experimental superscaling function, have been recently carried out both in the 
quasielastic and in the $\Delta$ regions and applied to neutrino reactions,
leading to results similar to the ones shown in Fig.~5.

\section{Conclusions}

We have shown how the full inelastic spectrum of inclusive electron-nucleus
scattering can be described within in a fully relativistic unified formalism
which can be applied in the few GeV energy domain.
The model, based on the Relativistic Fermi Gas, takes into account initial
and final state interactions in a phenomenological way through the use
of a superscaling function directly extracted from $(e,e')$ data. This allows
one to make model independent predictions for neutrino-nucleus cross sections,
needed in the analysis of $\nu$ oscillation experiments.

%
%


\end{document}